**The dark side of Open Access in Google and Google Scholar: the case of Latin-American repositories**

Enrique Orduña-Malea[1,*] and Emilio Delgado López-Cózar[2]

[1]EC3 Research Group, Universidad Politécnica de Valencia. Camino de Vera s/n, Valencia 46022, Spain.
[2]EC3 Research Group, Universidad de Granada, 18180 Granada, Spain
*e-mail: enorma@upv.es

**Abstract** Since repositories are a key tool in making scholarly knowledge open access (OA), determining their presence and impact on the Web is essential, particularly in Google (search engine par excellence) and Google Scholar (a tool increasingly used by researchers to search for academic information). The few studies conducted so far have been limited to very specific geographic areas (USA), which makes it necessary to find out what is happening in other regions that are not part of mainstream academia, and where repositories play a decisive role in the visibility of scholarly production. The main objective of this study is to ascertain the presence and visibility of Latin American repositories in Google and Google Scholar through the application of page count and visibility indicators. For a sample of 137 repositories, the results indicate that the indexing ratio is low in Google, and virtually nonexistent in Google Scholar; they also indicate a complete lack of correspondence between the repository records and the data produced by these two search tools. These results are mainly attributable to limitations arising from the use of description schemas that are incompatible with Google Scholar (repository design) and the reliability of web indicators (search engines). We conclude that neither Google nor Google Scholar accurately represent the actual size of open access content published by Latin American repositories; this may indicate a non-indexed, hidden side to OA, which could be limiting the dissemination and consumption of open access scholarly literature.

## 1. Introduction

Repositories, whose main functions were intended from the outset[1-2] to provide a deposit (facilitate self-archiving to preserve the academic legacy) and access facility (facilitate information retrieval processes), have become a key component of OA. According to *OpenDOAR* (as of December 2013), there are now over 2,500 repositories[3].

Among the various types of repositories in existence (Armbruster and Romary 2010), disciplinary (based on content produced in a specific area of knowledge) and institutional repositories (centred on documents produced by one or more institutions or organisations) play a key role in the direct dissemination of scholarly knowledge (Björk 2014; Chan 2004), although they have clearly distinct objectives and functions (Kling and McKim 2000).

While disciplinary repositories arise from a culture of informal communication between academics (a clear example of this is *arXiv*[4] or *RePEc*[5]) institutional repositories are designed to record the academic activity of one or more institutions (Ruiz-Conde and Calderón-Martínez 2014), and thus play a key role in the creation of the online digital identity of these organisations (Aguillo 2009).

Indeed, this conception of the institutional repository as deposit and service for disseminating and accessing its academic production (and thus a reflection of their activities), has led most universities around the world to create their own institutional repositories; the *Massachusetts Institute of Technology* (MIT)[6] was a pioneer in this field.

Thanks to the existence and the widespread use of both types of repository (disciplinary and institutional), the "green" route is now a reality (Björk et al. 2010). There is also empirical evidence showing that self-archiving, in which repositories play a central role, has already become the main driver behind OA. Archambault et al. (2013) detected, from a random sample of 160,000 articles published from 2008 to 2011 (retrieved from *Scopus*, *DOAJ* and *Pubmed central*), that approximately 68,610 documents (43%) are OA, of which 53,072 (77.3% of all OA articles) were published using the green route.

## 1.1. Visibility of the repositories on the Web and in Google search engines

Bearing in mind the importance of self-archiving for the generation of OA content, the presence and visibility of the repositories on the Web (and especially in search engines such as *Google* and *Google Scholar*) are essential to ensure that the content that they host makes OA truly effective for the community.

However, all this focus on the creation, design and dissemination of repositories generally takes the "Open Access" concept as its centre of gravity, forgetting that the product is, at the end of the day, a web site. As such, care should also be taken over all aspects relating to navigability, usability and visibility in search engines (Arlitsch et al. 2013). All these aspects are usually overlooked as they are almost completely entrusted to the software used to manage the collection, whose default setting is usually not the most suitable for website usability.

The main consequences of this are, first, the poor user experience of browsing and interacting with the repository and, second, the low indexing ratios for repository contents in search engines. This constitutes one of the main technical problems in the construction of repositories, due to its huge implications (Arlitsch and O'Brian 2012).

It is clear that correct indexing in *Google* (search engine par excellence), and *Google Scholar* (leading tool among researchers for finding scholarly information), is essential. It should be borne in mind that *Google* and *Google Scholar* have become the gateway to users searching for academic information. Already in 2005, the OCLC noted that 89% of students began their searches in search engines, and only 2% did so on the websites of academic libraries (DeRosa and *OCLC* 2005). Subsequent studies have corroborated this phenomenon. Griffiths and Brophy (2005) conducted user studies to gain a better understanding of student behaviour when using academic resources in the UK; they found that 45% of students used *Google* as their first option when locating information. Likewise, the *LibQual+ 2009 Survey* (Cook et al. 2009) showed that American students use search engines daily.

Haglund and Olsson (2008), meanwhile, conducted an observational study of researchers in Sweden from 2005 to 2006, in which they reveal how *Google* dominates as a starting point for researchers searching for scholarly information. These data were corroborated by the *Ithaka 2009* report (Schonfeld and Housewright 2010), in which a survey of 3,025 teachers was carried out in colleges and universities in the United States. The study concluded that researchers from the sciences, social sciences and humanities have gone from using library facilities and traditional search resources (catalogues and databases) to using search engines; and that both *Google* and *Google Scholar* count among the preferred methods for locating information in academic journals.

More recent reports, such as the *E-Expectations research reports*[7], also point out that users find information much more easily through *Google* than through the universities' own websites, which is a genuine handicap for the use and visibility of institutional repositories.

However, despite the changing informational behaviour of both students and researchers, the use of *Google* is instrumental in accessing scholarly information. Herrera (2011) analysed the *University of Mississippi*'s library logs*,* and revealed that the percentage of clicks originating from *Google Scholar* had grown from 4% in 2006 to 27% in 2009. This phenomenon is confirmed in Burns' doctoral thesis (2013), which demonstrated the predominance of the university when providing full text access to researchers using *Google Scholar* as their starting point in the search for scholarly information. Hence the clear importance of repositories having a presence in search engines, an aspect that is also key to the strategic future of university libraries.

Given the wide use of search engines as a starting point for queries – especially academic queries – and the fact that in most cases the resource is accessed directly without going through the repository website, it is vital that the contents of the repositories be properly indexed in search engines.

There have been few empirical studies on the extent to which repositories are indexed in search engines. These include the study by Arlitsch and O'Brian (2012), who found a very low indexing ratio for US institutional repositories in *Google Scholar*. Archambault et al. (2013) consulted various sources (*Google*, *Google Scholar* and *Microsoft Academic Research*) with the aim of locating OA versions of a sample of 500 articles published in 2008 and indexed in *Scopus*. They located an OA version of 48% of the sample (the search was conducted at the end of 2012). Furthermore, the authors found that if they only used *Google Scholar*, the percentage dropped to 41%. This means that much of the literature from the repositories is not retrievable in *Google Scholar*, even though the green route, as discussed above, is meant to give considerable weight to OA literature.

However, more studies on this subject are required, in order to extend research to other areas, especially those that are outside the academic mainstream. It is precisely in these areas that repositories play a decisive role in preventing academic endeavour from being lost; and this issue constitutes one of the main objectives of this study.

## 1.2. The cybermetric impact of repositories

The discipline of cybermetrics provides techniques and indicators suitable for analysing the extent to which repositories are indexed on search engines, as well as their web visibility and impact. However, although these techniques have generated considerable research output related to the academic world in general (Thelwall 2004) and to universities in particular (Orduña-Malea 2012), cybermetric analyses applied to repositories are still very scarce.

The cybermetric impact of repositories received a decisive boost with the *Ranking Web of Repositories*[8], developed in Spain by the *Cybermetrics Lab* under the aegis of the *Spanish National Research Council* (CSIC), and which is probably the project with the greatest reach and weight in the application of cybermetric indicators to repository analysis (Aguillo et al., 2010). This project positions repositories (both institutional and disciplinary) according to four web indicators: page count (measured by *Google*), visibility (*MajesticSEO*[9] and *Ahrefs*[10]), number of rich files (*Google*) and academic page count (*Google Scholar*).

Other studies have focused on the lack of precision in the cybermetric indicators owing to certain deficiencies in the following areas: the description of the repository metadata, as shown in the analysis of *OpenAIRE* (Aguillo 2011); the influence of URL syntax on the web visibility of repositories (Orduña-Malea and Regazzi 2014); and the analysis of the origins of hyperlinks to institutional repositories (Smith 2011; Sato and Itsumura 2011; Mas-Bleda et al. 2014).

For his part, Smith (2012; 2013) analysed the external links received by a set of Australian universities to detect any possible correlation with various bibliometric indicators related to the scholarly production of the institutions responsible for the repositories. In the first study, conducted through the *Blekko*[11] tool, Smith (2012) did not find any correlation between bibliometric and cybermetric indicators, suggesting two possible reasons for this: on the one hand, the existence of documents in the repositories that bear no direct relation to academic production, such as student work, graphic materials, institutional documents (minutes, reports), etc; and on the other hand, the variety of reasons, other than scholarly citation, for creating a hyperlink. This analysis was subsequently repeated, using *Google* instead of *Blekko* (Smith 2013), and achieved similar results, leading the author to conclude that the value of institutional repositories seems to be centred on making research accessible to the general web community rather than to the academic community in particular.

This argument again points to a failure in not approaching institutional repositories as websites (as well as bastions of academic knowledge). In this respect, consideration of certain indicators and metrics may help monitor the actual use of the repository and improve it, thus providing a more user-friendly experience to its real users.

Studies that have contributed to this field include those by Scholze (2007), who noted various methods to obtain repository usage data (*logs*), and Zuccala et al. (2007), who used both link and log file analysis to study the impact and use of an institutional repository. In another study (Zuccala et al. 2008), advanced link analysis techniques are used as a method to identify potential users or reveal hidden user communities, of strategic interest to repository managers since they provide predictable and consistent information.

In short, an analysis of repositories from the vantage point of cybermetrics may shed light on the web visibility and impact of content hosted on these platforms (particularly their indexing ratios in these search engines), essential for ensuring OA, as discussed above.

## 2. Objectives

The main objective of this study is to determine the visibility and impact of a representative sample of Latin American institutional repositories in order to:
- Find out the indexing ratios of these repositories in *Google* and *Google Scholar* in order to determine the possible extent of their invisibility.
- Apply web mention measures to ascertain the web impact of content published in repositories.
- Calculate the correlation between page count and impact indicators in order to determine whether there is a relationship between these dimensions in repositories.

## 3. Methodology

First, the process used to obtain the sample is outlined; then, the indicators used in the study are shown, together with the statistical treatment thereof.

### 3.1. Selection of the sample set

Selection of the sample set (Latin American repositories) was guided by an interest in analysing a set of highly cohesive repositories (in this case due to language and culture) hosted in countries that are in a development and growth cycle and that are outside the academic communications mainstream (i.e., underrepresented in the WoS and Scopus platforms). Their need for global scholarly dissemination and visibility is greater; hence repositories are an excellent medium for making their scholarly production accessible.

In order to obtain a significant sample, we selected all repositories listed in the *Ranking Web of Repositories* (July 2013 edition) under "Latin America", a total of 137 repositories (institutional in their vast majority). Full details of the repository names, corresponding URL and country are listed in the supplementary material[12]. The reason that this source was used instead of others (such as *OpenDOAR* or *ROAR*) is that it retrieves, in exhaustive and updated form, all existing repositories for which – given the URL syntax – analysis with web indicators is possible.

For each of the 137 repositories the URL was obtained and its syntax checked so that it could be used accurately in the cybermetric analysis (the use of subdomains within the domain of the repository's institution is recommended).

Table 1 shows the URLs for which incidents were detected, and indicates whether the URL was finally measured and under what conditions. As can be seen, most of the incidents are caused by using subdirectories instead of subdomains (although technically there is no difference between the two methods, search engine values for subdirectories are of limited accuracy).

Moreover, the automatic redirects that occur when accessing the resource create problems in other repositories (aspects that the webmaster should resolve through proper DNS management).

It should be noted that for URLs with a subdirectory, the subdirectory was eliminated in order to check whether there was a redirect. If there was, the base URL was considered;

if there was not, if an error was returned or another resource was accessed, then the analyses were performed taking the subdirectory into consideration. For example:
<site: repositorio.utp.edu.co/dspace>

**Table 1. Incidents in the syntax of the repository URLs**

| REPOSITORY (URL) | INCIDENT | MEASURED |
|---|---|---|
| intellectum.unisabana.edu.co:8080/jspui | Not accessible | Yes |
| repositorio.utp.edu.co/dspace | Subdirectory | Yes |
| uwispace.sta.uwi.edu/dspace | Subdirectory (DNS management) | Yes |
| bdigital.ces.edu.co:8080/dspace | Not located | No |
| repositorio.ufc.br | IP not located | No |
| bibliodigital.itcr.ac.cr/xmlui bibliodigital.itcr.ac.cr:8080/dspace | Subdirectory (DNS management; multidomain; no hierarchy) | Yes |
| tesis.udea.edu.co/dspace | Not located | No |
| repository.lasallista.edu.co/dspace | Subdirectory (DNS management) | Yes |
| cedes.ufsc.br:8080/xmlui | Subdirectory (base URL is another resource) | No |
| ru.ffyl.unam.mx:8080/jspui | Not located | No |
| repositorio.utfpr.edu.br/jspui/ | Subdirectory (DNS management) | Yes |
| repositorio.cti.gov.br/repositorio | Subdirectory (base URL is another resource) | Yes |
| repositorio.ufma.br:8080/jspui | Not located | No |
| repositorio.int.gov.br:8080/repositorio/ | Subdirectory (base URL is another resource) | No |
| campusesp.uchile.cl:8080/dspace/ | Subdirectory (DNS management) | No |
| acervo.ufvjm.edu.br:8080/jspui/ | Subdirectory (base URL is another resource) | No |
| repositorio.ub.edu.ar:8080/xmlui | DNS Management | No |
| repositorio.ehtc.cu/jspui | Subdirectory (DNS management; redirect errors) | Yes |
| biblio.colpos.mx:8080/jspui/ | Subdirectory (base URL is another resource) | No |

After the filtering process, the final sample consisted of a total of 127 URLs.

## 3.2. Web indicators

The different indicators used are shown in Table 2, indicating the source used to retrieve the information together with a brief definition.

**Table 2. Indicators, sources and definition**

| CATEGORY | INDICATOR | SOURCE | DEFINITION | QUERY |
|---|---|---|---|---|
| COUNT PAGE | Items (ITE) | Repository | Number of documents hosted by the repository | Direct method |
| | Total (Gtot) | Google | Number of files indexed on the website | site:domain.com |
| | PDF (Gpdf) | | Number of PDF files indexed | site:domain.com filetype:pdf |
| | Total Scholar (GStot) | Scholar | Number of files indexed on the website | site:domain.com |
| | PDF Scholar (GSpdf) | | Number of PDF files indexed on the website. | site:domain.com filetype:pdf |
| MENTION | URL mention (URL) | Google | Number of times the URL is mentioned. | "domain.com" – site:domain.com – inurl:domain.com |
| | Domains (V) | Open Site Explorer | Number of external links grouped by domain | Direct method |
| | MzRank (Mz) | | Link popularity score (0 to 10) | Direct method |

Measures of page count are intended to obtain data regarding, firstly, the actual size of the repository in number of items hosted (obtained from the information provided by the platform itself), and secondly, the total number of items listed in a search engine (in this case *Google* and *Google Scholar*). Additionally, data is retrieved for PDF files (both in *Google* and *Google Scholar*) as this format is commonly used for the final version of an academic product (Aguillo et al. 2010).

Moreover, mention values were obtained from the search engine *Open Site Explorer* (OSE).[13] This retrieves both the number of external links for each repository (measured at the aggregate domain level, i.e., all external links from the same domain are counted only once), and the *MzRank* indicator at subdomain level, which provides an estimated value for the popularity of the websites analysed, similar to *PageRank* (although for the selected sample, *MzRank* provides better results because it generates a scale of 1 to 100 instead of the *PageRank* scale of 1 to 10).

Additionally, the number of mentions for each URL was calculated from *Google*, which gave an estimated indicator of the number of external links (Ortega et al. 2014; Thelwall and Sud 2011), which is often used complementarily when the hyperlink source is not accessible or has insufficient coverage for the sample.

For each of the above, all the repository level indicators shown in Table 1 were manually applied. Subsequently, the data were transferred to a spreadsheet to be statistically analysed using the *XLStat* application, through which a correlation analysis was conducted for all indicators (given the unequal distribution of web data, the Spearman correlation coefficient was applied) as was a principal component analysis (PCA).

## 4. Results

### 4.1. Geographical Distribution

Table 3 shows the distribution by country of the 137 repositories analysed; the dominant countries are Brazil (37), Colombia (21), Argentina (18) and Ecuador (17).

**Table 3. Distribution of the repository sample by country**

| COUNTRY | N |
|---------|---|
| Brazil | 37 |
| Colombia | 21 |
| Argentina | 18 |
| Ecuador | 17 |
| Mexico | 12 |
| Chile | 8 |
| Venezuela | 8 |
| Peru | 5 |
| Costa Rica | 3 |
| Cuba | 3 |
| El Salvador | 2 |
| Jamaica | 2 |
| Puerto Rico | 1 |
| TOTAL | 137 |

In the case of Chile, it should be noted that one repository (*CONICYT Digital Repository*), has two URLs (<dspace.conicyt.cl/ri20> and <dspace2.conicyt.cl>), which for the purposes of this study have been treated independently.

### 4.2. Extent of repository indexing in *Google* and *Google Scholar*

Table 4 shows the overall data for the 20 repositories with the highest number of total items in their collections; *RedALyC Estudios Territoriales* prominently occupies first position (300,555 total items).

In addition to the number of items, total page count and PDF file data on *Google* and *Google Scholar* are displayed, as well as the percentages that these sizes represent of the total number of items in the repository (which is a way of indicating the indexing ratio of repository resources in the sources analysed)[14].

**Table 4. Page count indicators for repositories with the highest number of items**

| URL | ITEMS | PAGE COUNT | | | | | | | |
|---|---|---|---|---|---|---|---|---|---|
| | | GOOGLE* | | | | SCHOLAR | | | |
| | | PAGE COUNT | % | PDF | % | PAGE COUNT | % | PDF | % |
| estudiosterritoriales.org | 300,555 | 87,200 | 29.01 | 204 | 0.07 | 113 | 0.04 | 0 | 0.00 |
| lume.ufrgs.br | 75,986 | 134,000 | **176.35** | 58,000 | 76.33 | 42,100 | 55.40 | 191 | 0.25 |
| bibliotecadigital.icesi.edu.co | 68,017 | 256,000 | **376.38** | 6,000 | 8.82 | 28,500 | 41.90 | 270 | 0.40 |
| rad.unam.mx | 59,232 | 229,000 | **386.62** | 0 | 0.00 | 0 | 0.00 | 0 | 0.00 |
| bibliotecadigital.unicamp.br | 54,372 | 37,600 | 69.15 | 0 | 0.00 | 21,900 | 40.28 | 0 | 0.00 |
| dspace2.conicyt.cl | 49,173 | 6,860 | 13.95 | 156 | 0.32 | 89 | 0.18 | 0 | 0.00 |
| alice.cnptia.embrapa.br | 43,021 | 13,000 | 30.22 | 8,930 | 20.76 | 3,450 | 8.02 | 3,410 | 7.93 |
| saber.ucab.edu.ve | 42,534 | 524,000 | **1,231.96** | 11,600 | 27.27 | 6,680 | 15.71 | 0 | 0.00 |
| teses.usp.br | 42,243 | 299,000 | **707.81** | 36,800 | 87.12 | 1,630 | 3.86 | 140 | 0.33 |
| acervodigital.unesp.br | 40,409 | 13,000 | 32.17 | 584 | 1.45 | 158 | 0.39 | 3 | 0.01 |
| repositorio.ufsc.br | 33,691 | 72,500 | **215.19** | 17,400 | 51.65 | 30,000 | 89.04 | 158 | 0.47 |
| producao.usp.br | 29,169 | 53,000 | **181.70** | 5,000 | 17.14 | 93 | 0.32 | 1 | 0.00 |
| sedici.unlp.edu.ar | 28,512 | 556,000 | **1,950.06** | 14,500 | 50.86 | 20,000 | 70.15 | 532 | 1.87 |
| saber.ula.ve | 25,816 | 251,000 | **972.27** | 26,500 | 102.65 | 11,600 | 44.93 | 1,960 | 7.59 |
| dspace.espol.edu.ec | 23,292 | 557,000 | **2,391.38** | 23,800 | 102.18 | 11,400 | 48.94 | 19 | 0.08 |
| captura.uchile.cl | 20,832 | 468,000 | **2,246.54** | 6,060 | 29.09 | 13,300 | 63.84 | 18 | 0.09 |
| maxwell.lambda.ele.puc-rio.br | 17,352 | 24,200 | **139.47** | 13,900 | 80.11 | 652 | 3.76 | 573 | 3.30 |
| tesiuami.izt.uam.mx | 15,260 | 45,700 | **299.48** | 0 | 0.00 | 0 | 0.00 | 0 | 0.00 |
| cdigital.uv.mx | 13,895 | 174,000 | **1,252.25** | 18,900 | **136.02** | 6,570 | 47.28 | 40 | 0.29 |
| naturalis.fcnym.unlp.edu.ar | 13,782 | 184,000 | **1,335.07** | 1170 | 8.49 | 56 | 0.41 | 54 | 0.39 |

* Values for page count in Google that exceed the number of items are shown in bold

Errors in the functionality of search engines may be observed in Table 4 (full version in Annex II of the supplementary material), i.e., page count values for the repository lower than those shown for the search engines, although these errors vary according to the source.

In the case of *Google,* 109 URLs whose size is greater than the number of items were located. It therefore seems clear that the search engine is retrieving not only items from the repository but also other files hosted on the domain (including those pertaining to the application used to manage the repository). For PDF values, the number of URLs with this error is lower at 47, which indicates that this query is more accurate than that for overall size.

In the case of *Google Scholar,* there are even fewer errors. Total page count yields 11 URLs with page count values greater than those for the repositories, while for PDF files there are only three:

    <cybertesis.uach.cl> has 2,758 total PDF items and 2,900 in *Scholar* (105.15%).

<cybertesis.upc.edu.pe> has 114 total PDF items and 456 in *Scholar* (400%).
<ri.agro.uba.ar> has 82 total PDF items and 101 in *Scholar* (123.17%).

In this case, the errors are directly related to errors in the indexing of resources, but they are practically non-existent and are, in any case, detectable and easily controlled.

### 4.3. Impact of the repositories: mention indicators

Table 5 shows the 20 URLs with the best performance in URL mention, number of *referring domains* and *MzRank* indicators.

#### Table 5. Top 20 URLs for each mention indicator

| URL | URL Mention | URL | Referral domains | URL | MzRank |
|---|---|---|---|---|---|
| teses.usp.br | 5,380,000 | sedici.unlp.edu.ar | 194 | mord.mona.uwi.edu | 4.89 |
| repositorio.ufsc.br | 3,660,000 | dspace.c3sl.ufpr.br | 152 | sedici.unlp.edu.ar | 4.62 |
| producao.usp.br | 2,930,000 | rabci.org | 142 | bdigital.uncu.edu.ar | 4.46 |
| lume.ufrgs.br | 2,630,000 | bdigital.uncu.edu.ar | 139 | captura.uchile.cl | 4.44 |
| maxwell.lambda.ele.puc-rio.br | 1,240,000 | captura.uchile.cl | 83 | dspace.c3sl.ufpr.br | 4.37 |
| saber.ucab.edu.ve | 1,230,000 | rephip.unr.edu.ar | 74 | bvc.cgu.gov.br | 4.36 |
| ccdoc.iteso.mx | 1,050,000 | repository.urosario.edu.co | 68 | cybertesis.upc.edu.pe | 4.35 |
| bdm.bce.unb.br | 869,000 | bdm.bce.unb.br | 67 | bibdigital.epn.edu.ec | 4.29 |
| dspace.c3sl.ufpr.br | 829,000 | cybertesis.uach.cl | 67 | repositorio.usfq.edu.ec | 4.27 |
| saber.ula.ve | 778,000 | cdigital.uv.mx | 62 | repositorio.espe.edu.ec | 4.24 |
| bibliotecadigital.unicamp.br | 701,000 | saber.ula.ve | 61 | tesis.pucp.edu.pe | 4.22 |
| sedici.unlp.edu.ar | 672,000 | repositorio.espe.edu.ec | 60 | producao.usp.br | 4.19 |
| repositorio.unb.br | 601,000 | repositorio.uasb.edu.ec | 60 | cybertesis.uach.cl | 4.18 |
| alice.cnptia.embrapa.br | 583,000 | bibdigital.epn.edu.ec | 59 | rad.unam.mx | 4.18 |
| eprints.uanl.mx | 510,000 | dspace.ups.edu.ec | 58 | naturalis.fcnym.unlp.edu.ar | 4.16 |
| tesis.uchile.cl | 447,000 | saber.ucv.ve | 57 | digital.bl.fcen.uba.ar | 4.15 |
| cybertesis.uach.cl | 373,000 | mord.mona.uwi.edu | 57 | repositorio.utn.edu.ec | 4.15 |
| dspace.ups.edu.ec | 366,000 | tesis.pucp.edu.pe | 55 | repositorio.iaen.edu.ec | 4.15 |
| ri.biblioteca.udo.edu.ve | 322,000 | digital.bl.fcen.uba.ar | 54 | rephip.unr.edu.ar | 4.14 |
| bibliotecadigital.uel.br | 322,000 | dspace.espoch.edu.ec | 54 | dspace.espoch.edu.ec | 4.13 |

In the case of URL mentions, the values obtained are exceptionally high, especially for <tesis.usp.br> (5,380,000 hits). Although search engines round up these values, it is evident that extra noise is high, despite using the <-inurl> command to exclude certain types of *spam*. Even so, we detected some exceptions in some URLs, which, despite having high page count values (for items both in the repository and indexed by *Google*), made hardly any impact in URL mentions. This is the case of, for example, <repositorio.uasb.edu.ec> (5,370 mentions) and <uwispace.sta.uwi.edu/dspace> (216 mentions), although their positions in the Web Ranking are relatively high.

In the case of the *referring domains*, the achieved impact was very low: only 4 URLs achieved more than 100 domains, while 21 did not return any results. These data correspond to the *MzRank* values (which depend directly on the quantity and quality of inbound external links on the analysed websites). In this case, and as shown in Table 5, no URL scored more than 5 points (the maximum is 10). Moreover, 23 URLs obtained a "0" value (sometimes OSE covers the subdomain corresponding to the repository).

**4.4. Correlation between page count and impact**

Correlations between all the web indicators (page count and mention) are shown in Table 6. As can be observed, the number of items retrieved directly from the platform (ITE) correlated significantly with various mention indicators, especially with PDF file page count in *Google* (r=.75) and total page count in *Scholar* (r=.68). However, a very low correlation was obtained with PDF page count in *Google Scholar* (r=.31), when it was precisely this indicator which should have been the most accurate in capturing the number of articles deposited in an institutional repository; it returned very low indexing ratios, as could already be observed in Table 4 above.

With regard to the correlation of ITE with mention indicators, unexpectedly significant results were achieved with the number of URL mentions (r=.63), which demonstrates that despite the document noise of this indicator, the results do have certain value.

Finally, almost no correlation was observed between ITE and indicators related to hyperlinks, both for the number of *referring domains* (r=.26) and for *MzRank* (r=.22).

**Table 6. Correlation between indicators**

|        | ITE    | Gtot   | Gpdf   | GStot  | GSpdf  | URL    | V      | Mz  |
|--------|--------|--------|--------|--------|--------|--------|--------|-----|
| ITE    | 1      |        |        |        |        |        |        |     |
| Gtot   | 0.592* | 1      |        |        |        |        |        |     |
| Gpdf   | 0.752* | 0.730* | 1      |        |        |        |        |     |
| GStot  | 0.683* | 0.642* | 0.795* | 1      |        |        |        |     |
| GSpdf  | 0.315* | 0.189  | 0.472* | 0.357* | 1      |        |        |     |
| URL    | 0.639* | 0.329* | 0.589* | 0.534* | 0.444* | 1      |        |     |
| V      | 0.265* | 0.303* | 0.373* | 0.396* | 0.299* | 0.383* | 1      |     |
| Mz     | 0.227  | 0.358* | 0.284* | 0.236* | 0.182  | 0.364* | 0.768* | 1   |

*\* Significant values (except diagonal) at the level of significance alpha=0.050 (two-tailed test)*
ITE: number of items; Gtot: number of files in Google; Gpdf: number of PDF files in Google; GStot: number of files in Scholar; GSpdf: number of PDF files in Scholar; URL: number of URL mentions in Google; V: number of domains referring to; Mz: Mzrank value.

To complement these data we conducted a Principal Component Analysis (PCA), shown in Figure 1.

The PCA clearly shows the separation between performance in page count and visibility, and how the URL mention indicator seems closer to the page count than to the visibility indicators, when by their nature the opposite should be true.

**Figure 1. Principal Component Analysis (PCA) of Latin American repositories**

**5. Discussion and conclusions**

The main results for the indexing ratios of repository contents in search engines, their web impact and the relationship between web page count and visibility are discussed below.

*Indexing ratios*
Before commenting on the repository document indexing ratio in *Google* and *Google Scholar*, the complete lack of correspondence between the repository records and the data produced by these two tools should be noted. Equally striking are the highly marked discrepancies in information among the search engines themselves: they only coincide in their extremely low indexing values for PDF documents.

This raises a preliminary question about the reliability and validity of the data search and recovery process ("site" command), the technical indexing mechanisms of the robots used by *Google* and *Google Scholar* and/or the deficient web architecture of the repositories themselves, which could well be the cause that lies behind the other aspects. Similarly, the design of the database of some of the repositories may prevent the accurate retrieval of indicators by search engine robots (a concept known as the invisible internet), although the development of applications such as DSpace (widely used in the installation of this study's sample repositories) has eliminated this problem.

With regard to *Google* (which should, in principle, index everything to achieve its goal of making the world's information universally accessible), the inordinately high page count data (well above real values) must be due to the counting of files that are not specifically items of the collection studied, i.e., files pertaining to the software itself or other information hosted by the server being analysed (easily verifiable by manually browsing through the results returned for the "site" query in the search engine).

Regarding the number of PDF documents, although exact figures for this document type in the repositories under study are not known, such a low indexing ratio is very strange. The pervasive use of the PDF format is an irrefutable fact in academia (Aguillo 2009), and it is very odd that academic repositories such as those studied here, which often contain scholarly output – theses, articles, reports and other academic documents (course programmes, teaching materials) – have such a low percentage, save a few notable exceptions (<lume.ufrgs.br>, <repositorio.ufsc.br>). It is therefore plausible to conclude that *Google* underrepresents the scientific and academic content of the repositories.

By contrast, the total number of documents indexed in *Google Scholar*, contrary to *Google*, is well below what was expected. The low item indexing ratios in *Google Scholar* (whose database is not the same as *Google*'s) are consistent with those obtained previously by Arlitsch and O'Brian (2012), who detected low indexing ratios in the United States for repository articles in *Google Scholar,* and with data from the *Ranking Web of Repositories*.

In any case, there are several reasons that may explain why the overall data should be viewed with some caution. First, because the "site" operator does not return all the items that *Google Scholar* has indexed for a repository, which means it is not exhaustive. Second, because the system of grouping multiple versions of an article operates in such a way that one version is taken as the "primary" version. This process is done automatically[15], although authors may also manually select which is the main version of the article. The "site" command only returns data for the main version. This means that if an article is hosted on different platforms (e.g. journal and repository), and if the primary version is the one published in the journal, the "site" operator applied to the repository will not count the item and vice versa, although it is indexed on both platforms.

This particularly affects the accuracy of *Google Scholar* in measuring the performance of repositories with this indicator, and largely explains the low values. It also opens up a future line of work which should consider whether the repositories with better indexing ratios in Google *Scholar* are also those with higher numbers of primary versions amongst their items, which may explain the better results of some repositories compared to others.

What does stand to reason is that *Google Scholar* indexes far fewer PDF documents than *Google,* given the requirements and recommendations that this search engine provides institutional repository webmasters for indexing documents. These include the following:

> "If you're a university repository, we recommend that you use the latest version of Eprints (eprints.org), Digital Commons (digitalcommons.bepress.com), or DSpace (dspace.org) software to host your papers."

Arlitsch and O'Brian (2012), while noting the limitations of the "site" command, found that the main causes are the metadata schema used and the navigability and information architecture features, which do not help the search engine robots carry out the indexing processes correctly. Indeed, they applied various changes to the description schema (rejecting *Dublin Core* in favour of other schemas recommended by *Google Scholar*, such as *Highwire Press*), and indexing ratios improved significantly over time. These limitations of *Google Scholar* in measuring the presence of repository contents also contrast with the policy of certain products of this company, such as *Google Scholar Metrics*, which quantifies the scholarly impact of repositories (Delgado López-Cózar and Robinson-García 2012).

In short, it may be concluded that the low repository content indexing ratios are mainly due to these two limitations: the use of description schemas that are not compatible with *Google Scholar* (repository design) and the reliability of the web indicators (search engines).

Finally, it was found that the queries that combined the overall page count with the PDF file type in *Google* were those that achieved the optimal results, and that were most similar to the data that the repositories themselves indicated with regard to the size of their collections. This may have been determined by the fact that primary versions are not accounted for in the search – whereas in *Scholar* they are – which clearly underrepresents the presence of repositories when measured by the "site" command.

***Impact values and their relationship to web page count***
The mention values generally exhibited excessive noise for the URL mention indicator, as well as relatively low values both for the number of external links and for the *MzRank* indicator.

Moreover, the correlation between page count and visibility indicators is low. The highest value is between URL mentions and PDF file page count in *Scholar* (r=.589), which we consider to be lower than expected, especially considering the noise in the URL mention indicator, which may have caused a slight increase in this correlation.

Scholarly literature has previously shown how these two indicators generally have a very significant correlation since, in statistical terms, the websites with the greatest number of pages (page count) are more likely to receive more links (visibility). This has been amply demonstrated in academic environments (Aguillo & Granadino 2006), when the university is measured as a whole. This is the main reason that the web impact factor (WIF) is an indicator that is not used (large or small websites can achieve the same WIF).

However, Orduña-Malea (2013), in a study of the Spanish university system, demonstrated how, on disaggregated levels of the university (including institutional repositories), this relationship is not satisfied, thus reflecting a lack of visibility in comparison with the existing page count.

Therefore, the low correlation indicates and clearly confirms that at the "repository" level there is a performance gap between page count and impact. This means that the real repository collections are beginning to grow, but their web visibility (measured by the number of mentions and links they receive) is still very low. Low content indexing ratios, which cause the measured page count to be smaller than it really is, compound the low visibility values. These results are consistent with those obtained previously by Orduña-Malea and Regazzi (2014), who detected this situation in the US university system, which confirms that it is not an effect that pertains specifically to the region studied (Latin America).

Finally, with respect to the correlation between the various web visibility indicators, a strong correlation between V and Mz (.768) was observed, although there was very little correlation between these two measurements and URL mentions (.383 and .364 respectively). The PCA analysis confirms visually how URL mentions seem to be closer to page count values. This unusual effect is ascribed to the excessive document noise for this indicator, the use of which is not recommended in cybermetric analysis of repositories.

### Final conclusions

The results of this study highlight the insufficient dissemination of open access scholarly literature (crucially in terms of web visibility) in a medium (the Web) that is by definition its natural environment, and in a context (Latin America), in which scholarly production requires extra visibility because it lies outside the academic mainstream (i.e. not published in journals indexed in WoS or Scopus).

Given the weight of the green route in the dissemination of OA scholarly literature, and the importance of *Google* (and *Google Scholar*) to the search and use of academic information, the low visibility of the contents could significantly affect the real use of OA by end users. It would appear to be generating a great hidden mass of open access content, from institutional repositories, which neither Google, in the first instance, or users, in the last instance, can locate.

The lack of web visibility of the analysed repositories is determined by the low indexing ratios of their content (both in *Google* and *Google Scholar*), since a low web presence determines a corresponding low web visibility.

These low indexing ratios are, in turn, determined by the use of description schemas that are ill-suited to *Google* and inadequate web navigability, factors already outlined by Arlitsch and O'Brien (2012). Additionally, this study has also identified certain technical limitations in the use of web indicators in *Google* and *Google Scholar* to measure this indexing.

Therefore, we consider that neither *Google* nor *Google Scholar* are accurate or representative of the actual page count of open access content published by Latin American repositories; this may indicate the existence of a hidden, non-indexed side of OA.

In any case, the technical limitations of *Google Scholar*, in only counting primary versions of articles, tilt the balance towards the use of *Google* to measure page count, despite the fact that the document noise is greater. However, a thorough analysis of the real influence of the primary version search and accuracy of the "site" command in repository performance in *Google Scholar* (which requires an item by item analysis of each collection) is deemed necessary.

Much of the solution to these problems is purely technical, and should be addressed in the short term to ensure the visibility of repositories, to which institutions are devoting significant financial and human resources. This must include a rethinking of the goals that must be achieved to guarantee the success of a repository, for which presence and visibility in search engines must bear greater weight.

However, the results come from the analysis of a small sample of repositories, and should be widened in the future to larger samples in order to draw more definitive conclusions.

## 6. Notes

[1] http://www.budapestopenaccessinitiative.org (accessed 15 March 2014).
[2] http://oa.mpg.de/lang/en-uk/berlin-prozess/berliner-erklarung (accessed 15 March 2014).
[3] http://www.opendoar.org/index.html (accessed 15 March 2014).
[4] http://arxiv.org (accessed 15 March 2014).
[5] http://repec.org (accessed 15 March 2014).
[6] http://dspace.mit.edu (accessed 15 March 2014).
[7] E-Expectations research reports.
http://omniupdate.com/resources/research.html (accessed 15 March 2014).
[8] http://repositories.webometrics.info (accessed 15 March 2014).
[9] http://www.majesticseo.com (accessed 15 March 2014).
[10] http://ahrefs.com (accessed 15 March 2014).
[11] http://blekko.com (accessed 15 March 2014).
[12] http://hdl.handle.net/10481/32271 (accessed 17 June 2014)
[13] http://www.opensiteexplorer.org (accessed 15 March 2014).
[14] Of the 127 URLs analysed, page count data were not obtained for five, which have not been taken into account for the rest of the calculations: <bibliotecavirtual.unl.edu.ar>, <memoria.fahce.unlp.edu.ar>, <repositorio.utm.edu.ec>, <dspace.conicyt.cl/ri20> y <cartapacio.edu.ar>.
[15] http://www.google.com/patents/US8589784 (accessed 15 March 2014).